\documentclass[amsmath,amssymb,11pt]{revtex4}
\usepackage{psfig}
\usepackage{graphicx}
\usepackage{dcolumn}
\usepackage{bm}
\usepackage{amsmath}
\usepackage{amsfonts}
\usepackage{amssymb}%
\newcommand{\n}{\ensuremath{\nu(t,r)}}
\newcommand{\T}{\ensuremath{\theta}}
\newcommand{\pt}{\ensuremath{p_{\theta}}}
\newcommand{\M}{\ensuremath{{\cal M}}}

\newcommand{\X}{\ensuremath{{\cal X}}}
\begin{document}
\preprint{}

\title{Gravitational Collapse End States}
\author{{ Pankaj S Joshi}}
\affiliation{ Department of Astronomy and Astrophysics\\ Tata
Institute of Fundamental Research\\ Homi Bhabha Road,
Mumbai 400 005, India}

\begin{abstract} Recent developments on the final state of
a gravitationally collapsing massive matter cloud are summarized
and reviewed here. After a brief background on the problem,
we point out how the black hole and naked singularity end states
arise naturally in spherical collapse. We see that it is the geometry 
of trapped surfaces that governs this phenomena. 
\end{abstract}

\maketitle

It was pointed out by S. Chandrasekhar in 1935:

"...The life history of
a star of small mass must be essentially different from that of a star of
large mass... A small mass star passes into White-dwarf stage... A star of
large mass {\it cannot} pass into this stage and one is left speculating
on other possibilities."

The question that {\it what happens when a star dies} has been a key
problem in astronomy and astrophysics for past decades.
If the star is sufficiently
massive, beyond the white dwarf or neutron star mass limits, then a 
continued gravitational collapse must ensue when the star has exhausted 
its nuclear fuel.\\

What are the  {\it possible end states}  of such a {\it continued 
gravitational collapse}?
To answer this question, one must study {\it dynamical 
collapse scenarios} within the framework of a gravitation theory 
such as Einstein's theory.  I shall try to give a somewhat 
non-technical account of recent developments in this area in recent years.

Penrose conjectured in 1969, that the ultra-dense regions i.e.
the spacetime singularities 
(where the physical quantities  e.g. densities, curvatures are having 
extreme values) forming 
in gravitational collapse must be hidden within the event horizon of 
gravity, that is, 
the collapse must end in a black hole. 
This is called the Cosmic Censorship Conjecture. There is
however no proof, or any suitable mathematical formulation of the same, 
available as of today.
\par 

Under the situation, very many researchers have made extensive study of 
various dynamical collapse models, mainly spherically symmetric, 
over past decade and a half, to
investigate the final outcome of an endless gravitational collapse.
When no proof, or even a suitable mathematical formulation
of censorship conjecture is available, it is only such studies that
can throw light on this issue. 
The generic conclusion appears to be: Either a {Black Hole}(BH) 
or a {Naked Singularity}(NS)
develops as end product of collapse, depending on the initial
data (e.g. initial density, pressures, and velocity profiles for the
collapsing shells) from which the
collapse develops, and the nature of dynamical evolutions
as permitted by Einstein equations.

While extensive study is made of astrophysics
of Black Holes, for NS we may still want to inquire into questions 
such as, {\it Are Naked Singularities Generic}, or {\it What are the
Physical factors} that cause NS, rather than a BH forming as 
Collapse End State? That is, one may wish to understand 
in a better way the
naked singularity formation in gravitational collapse. 
We try to address these issues in some detail here, 
and this may also possibly give an idea on the present status of
Cosmic Censorship Conjecture. For some further details and
references, we refer to [1-6].

A spherically symmetric spacetime with the general metric in 
{\it t, r, $\theta$, $\phi$} co-ordinates can be written as,
\begin{equation}
ds^2=-e^{2\nu(r,t)}dt^2+e^{2\psi(r,t)}dr^2+R^2(t,r)d\Omega^2 
\label{eq:metric}
\end{equation}
In such a comoving reference frame, the energy-momentum tensor  for a 
general {\it `Type I'} matter field can be written as, 
\begin{equation}
T^t_t=-\rho;\; T^r_r= p_r;\; T^\T_\T=T^\phi_\phi= p_\T 
\label{eq:setensor}
\end{equation}
This includes practically all physically reasonable matter forms such as 
dust, perfect fluids, massless scalar fields, and such other forms of matter.

We always consider the matter to satisfy a suitable energy condition,
Such as the {\it  Weak Energy Condition}:
\begin{equation}
T_{ik}V^iV^k\ge0 
\end{equation}
The point here is that, while it may be possible that energy condition
violations may occur in the very late stages of collapse when the quantum
effects take over, they should necessarily be satisfied for the classical
phase which governs most of the collapse evolution.

The {\it Einstein equations} then have the following form:
\begin{eqnarray}
\rho=\frac{F'}{R^2R'}; && p_r=\frac{-\dot{F}}{R^2\dot{R}}
\label{eq:ein1}
\end{eqnarray}
\begin{equation}
\nu'=\frac{2(\pt-p_r)}{\rho+p_r}\frac{R'}{R}-\frac{p_r'}{\rho+p_r}
\label{eq:ein2}
\end{equation}
\begin{equation}
-2\dot{R}'+R'\frac{\dot{G}}{G}+\dot{R}\frac{H'}{H}=0
\label{eq:ein3}
\end{equation}
\begin{equation}
G-H=1-\frac{F}{R}\nonumber
\label{eq:ein4}
\end{equation}

The functions $G$ and $H$ above are defined as,
\begin{equation}
G(t,r)=e^{-2\psi}(R')^2, H(t,r)=e^{-2\nu}(\dot{R})^2
\end{equation}
Also $F=F(t,r)$ is an arbitrary function, and has the 
interpretation of the mass function of the collapsing cloud, with  
$F\ge0$. Various {\it Regularity Conditions} ensure that the 
collapse develops from a Regular Initial Data.
From the point of view of dynamic evolution of initial data, at 
the initial epoch $t=t_i$, 
we have six arbitrary functions of the radial coordinate $r$,
\begin{eqnarray}
\nu(t_i,r)=\nu_0(r); & \psi(t_i,r)=\psi_0(r); & R(t_i,r)=R_0(r) \nonumber \\
\rho(t_i,r)=\rho_0(r); & p_r(t_i,r)=p_{r_0}(r); & \pt(t_i,r)=p_{\T_0}(r)
\label{eq:init}
\end{eqnarray}
All the initial data represented above are not mutually independent, as from 
{\it Bianchi Identity} we have,
\begin{equation}
\nu_0(r)=\int_0^r\left(\frac{2(p_{\T_0}-p_{r_0})}
{r(\rho_0+p_{r_0})}-\frac{p_{r_0}'}{\rho_0+p_{r_0}}\right)dr\nonumber 
\label{eq:nu0}
\end{equation}

Hence we see that there are five total field equations with seven 
unknowns, { $\rho$}, { $p_r$}, { $p_\theta$}, {$\psi$}, 
{ $\nu$}, { $R$}, and { $F$}, giving us 
the freedom of choice of two free functions. Selection of these 
functions, subject to the weak energy condition, and the given initial 
data for collapse at the initial spacelike surface, 
determines the matter distribution and metric of the space-time, and 
thus leads to a particular time evolution of collapse from the given
initial data.
Let us now consider a general mass function $F(t,r)$ for the 
collapsing
cloud, given as
\begin{equation}
F(t,r)=r^3\M (r,v)
\end{equation}
where {$\M$} is any regular and suitably differentiable
function.
We note here that we are NOT making any choice regarding the mass function,
as it is in the most general form.

Using the freedom of choice mentioned above, let us choose,
\begin{equation}
\n=A(t,R)
\end{equation}
(assuming that there are no shell-crossings in the spacetime, one 
could also make a more general choice as defined by
$\n'= A(r,v)_{v} R'$). 
Then we get the solution to {\it Einstein equation} as, 
\begin{equation}
G(t,r)=b(r)e^{2(A-\int A_{,t}dt)}
\end{equation}
Here {$b(r)$} is another arbitrary function of $r$.
In correspondence with dust models, we write, 
\begin{equation}
b(r)=1+r^2b_0(r)
\end{equation}
where {$b_0(r)$} is the velocity distribution function of the
collapsing shells. The marginally bound case corresponds to 
$b_0(r)=0$.

It is now possible to go back to the Einstein equations 
and work out the function { $t(v,r)$}. The physical spacetime singularity
develops at the value 
{$R=0$}, which corresponds to the physical radius of all collapsing shells 
going to zero. This corresponds to the time  {$t_s(r)=t(0,r)$}.
At the initial epoch we have $v=1$ and the singularity is defined
by the value $v=0$. 
Expanding  the function $t(v,r)$ around the center, we get,
\begin{equation}
t(v,r)=t(v,0)+r\X(v)+O(r^2)
\end{equation}
We then get the tangent to the singularity curve at the central 
singularity as,
\begin{equation}
\X(0)=-\frac{1}{2}\int_0^1dv\frac{\sqrt{v}(b_1v+vh_1(v)+\M_1(v))}
{\left(b_{00}v+vh_0(v)+\M_0(v)\right)^{\frac{3}{2}}}
\end{equation}
To decide the final fate of collapse in terms of either
a Black Hole or a Naked Singularity, we need to study the 
behavior of the apparent
horizon in the spacetime, and to examine if there are any families of outgoing
non-spacelike trajectories, which terminate in the past at the 
singularity.

The {\it apparent horizon} within the collapsing cloud is given by 
{$R/F=1$}, which gives the boundary of the trapped surface region 
of the space-time. If the neighborhood of the center gets trapped 
earlier than the singularity, then it will be covered. Otherwise, 
the singularity could be naked with non-spacelike future directed 
trajectories escaping from it to outside observers.

In order to consider the families of outgoing null geodesics from 
the singularity, 
and to examine the nature of the central singularity at 
$R=0,\;\; r=0$, 
let us consider the outgoing radial null geodesics equation,
\begin{equation}
\frac{dt}{dr}=e^{\psi-\nu}
\label{eq:null1}
\end{equation}
The singularity occurs at {$v(t_s(r),r)=0$}. 
Therefore, if there are any future directed null geodesics 
terminating in the past at the singularity, we must have 
{$R\rightarrow0$} as {$t\rightarrow t_s$} 
along these curves.
Rewriting the null geodesic equation in terms of 
the variables 
{$(u=r^\alpha,R)$}, it can be further analyzed,
\begin{equation}
\frac{dR}{du}=\frac{1}{\alpha}r^{-(\alpha-1)}R'\left
[1+\frac{\dot{R}}{R'}e^{\psi-\nu}\right]
\end{equation}
Choosing {$\alpha=\frac{5}{3}$} we get
\begin{equation}
\frac{dR}{du}=\frac{3}{5}\left(\frac{R}{u}+\frac{\sqrt{v}v'}
{\sqrt{\frac{R}{u}}}\right)\left(\frac{1-\frac{F}{R}}
{\sqrt{G}[\sqrt{G}+\sqrt{H}]}\right)
\end{equation}
The tangent to the null geodesics from the 
singularity is,
\begin{equation}
x_0=\lim_{t\rightarrow t_s}\lim_{r\rightarrow 0} \frac{R}{u}
=\left.\frac{dR}{du}\right|_{t\rightarrow t_s;r\rightarrow 0}
\end{equation}
Using the null geodesic equation the value $x_0$ is given as,
\begin{equation}
x_0^{\frac{3}{2}}=\frac{5}{\sqrt{6}}\X(0)
\end{equation}
Then, whenever $\X(0)>0$, there is radial null 
geodesic coming out from the singularity as given by  
$t-t_s(0)=x_0r^{\frac{5}{3}}$.

It has been already stated that we have the choice of two free 
functions in the above analysis, among which we are choosing only 
one so far,
in the most general case, namely, the form of metric function {$\n$}.
Therefore, the analysis holds for any equation of state of the form
\begin{equation}
p_r=Q(\rho)
\end{equation}
In this case there would be a constraint on otherwise arbitrary 
function {$\M$}. 
Furthermore the value of the function {$p_\theta(t,r)$} in terms of 
{$\rho(t,r)$} will be determined from Einstein equations.

In other words, it turns out that the BH/NS phases are determined by the
geometry of the trapped surfaces and the apparent horizon
that develop as the collapse evolves.

What governs the geometry of the trapped surfaces, or the 
formation or otherwise of the naked singularities in the spacetime?
In other words, what is it that causes the naked singularity to
develop rather than a black hole as the final end product of
collapse?
It turns out that physical agencies such as 
INHOMOGENEITIES in matter profiles as well as spacetime SHEAR 
play an important role to distort the trapped surface geometry,
and could delay the trapped surface formation during the collapse, thus
giving rise to NS, rather than a BH, as collapse end state
(see e.g. references [7-11]).

This, in a way, provides the physical understanding
of the phenomena of BH/NS end states.

It is seen that in spherical collapse the BH/NS phases as end state of
collapse are generic, and are seen to be determined by the
nature of the initial data from which the collapse develops,
and in terms of the allowed dynamical evolutions.
Physical agencies such as Inhomogeneities and Shear
cause them, and given the initial data, there are non-zero measure
classes of evolutions
that evolve into either of these outcomes.

What does the Non-spherical Collapse do? There are
some examples which indicate outcome to be similar ---but the 
evidence in that case is somewhat limited so far. The main difficulty
is the complexity of the Einstein system, which is second order,
non-linear partial differential equations.

There are several quite interesting questions which are under
active investigation at the moment. For example, could naked
singularities generate bursts of  Gravity Waves? What kind of quantum
effects will take place near NS? What will be the generic outcome
for the case of non-spherical collapse? Many of these issues
would have interesting physical implications.
It appears quite likely from the current investigations that the astrophysical 
phenomena such
as the Gamma Rays Bursts will have a strong connection to
the physics and dynamics of gravitational collapse of massive stars. 
Another intriguing possibility could be, NS may possibly provide us with 
some kind of observational signatures for the Quantum Gravity 
effects from ultra-strong gravity regions. This last possibility
would then be considered quite an exciting prospect.

\end{document}